\documentclass[twoside]{article}

\usepackage[accepted]{aistats2022}
\usepackage[round]{natbib}
\usepackage{xurl}

\usepackage{subcaption}
\usepackage{amsfonts}
\usepackage[ruled,vlined]{algorithm2e}
\usepackage{graphicx}
\usepackage{arydshln}
\usepackage{anyfontsize}
\usepackage{wrapfig}

\DeclareMathAlphabet{\mathcal}{OMS}{cmsy}{m}{n}

\def\mcz{\mathcal{Z}}

\def\tbfz{\tilde{\mathbf{z}}}
\def\bfZ{\mathbf{Z}}
\def\bfU{\mathbf{U}}

\def\tbfZ{\tilde{\mathbf{Z}}}
\def\twZ{\bfZ_{(t-W,t]}}

\begin{document}

%

%

\twocolumn[

\aistatstitle{Sequential Multivariate Change Detection with Calibrated and Memoryless False Detection Rates}

\aistatsauthor{ Oliver Cobb \And Arnaud Van Looveren \And  Janis Klaise }

\aistatsaddress{ Seldon Technologies \And Seldon Technologies \And Seldon Technologies} ]

\begin{abstract}
Responding appropriately to the detections of a sequential change detector requires knowledge of the rate at which false positives occur in the absence of change. Setting detection thresholds to achieve a desired false positive rate is challenging. Existing works resort to setting time-invariant thresholds that focus on the expected runtime of the detector in the absence of change, either bounding it loosely from below or targeting it directly but with asymptotic arguments that we show cause significant miscalibration in practice. We present a simulation-based approach to setting time-varying thresholds that allows a desired expected runtime to be accurately targeted whilst additionally keeping the false positive rate constant across time steps. Whilst the approach to threshold setting is metric agnostic, we show how the cost of using the popular quadratic time MMD estimator can be reduced from $O(N^2B)$ to $O(N^2+NB)$ during configuration and from $O(N^2)$ to $O(N)$ during operation, where $N$ and $B$ are the numbers of reference and bootstrap samples respectively.
\end{abstract}

\section{INTRODUCTION}

Algorithms that can detect change in the distribution underlying a data stream have long been important in applications such as quality assurance and cybersecurity. However there is a growing need for algorithms addressing the more specific problem of detecting when the distribution underlying a stream changes from that which generated a historical reference set.

\begin{figure}[t]
    \centering
    \includegraphics[width=\linewidth, trim={2mm, 5mm, 5mm, 0mm}, clip]{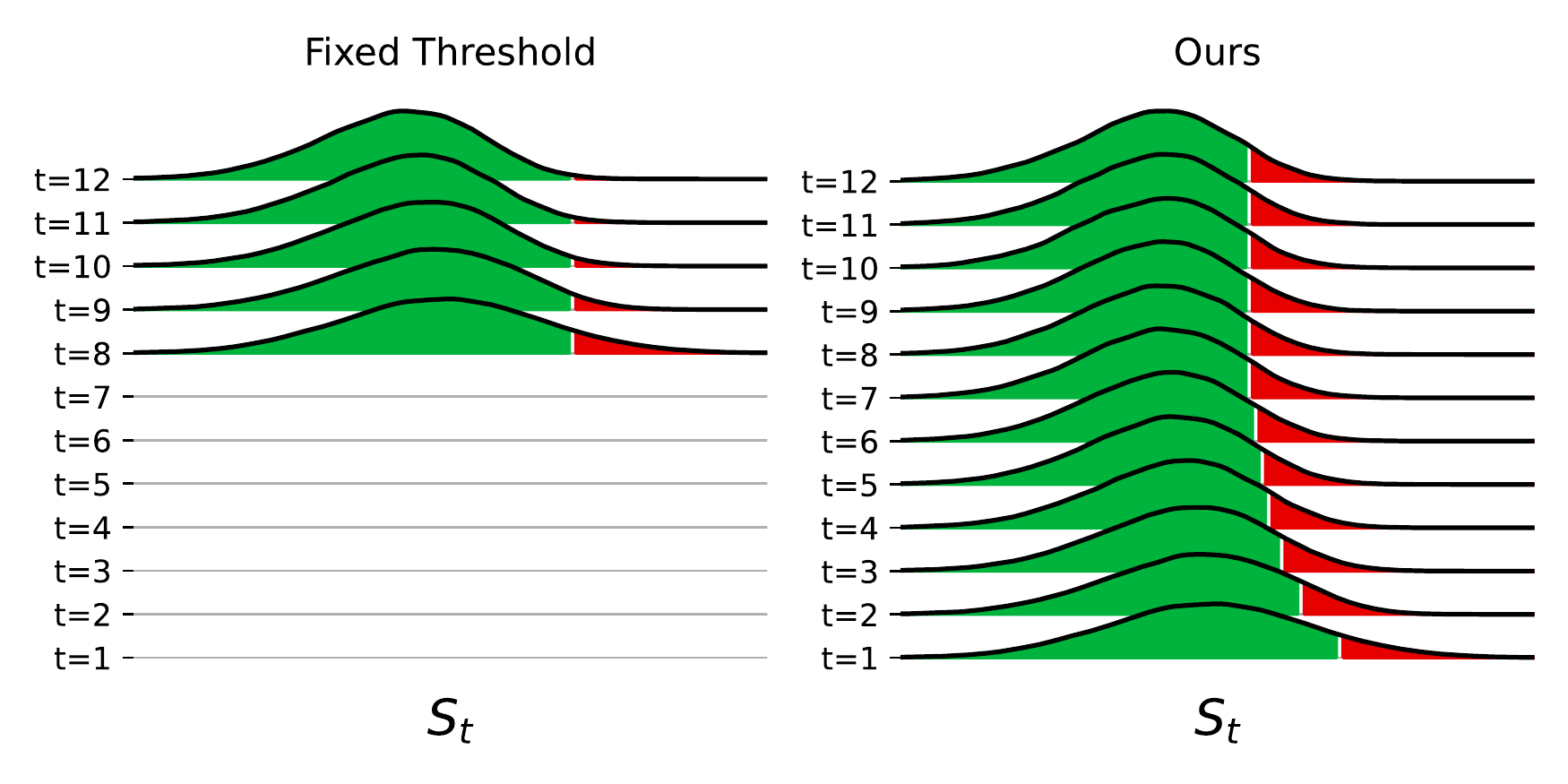}
    \caption{When using a sliding window of size $W$, existing approaches typically perform the first test at time $t=W$ with a threshold chosen such that it is exceeded by the first test-statistic $S_W$ with false-positive probability $\alpha$. Subsequent test-statistics $S_t$ are then compared to the same threshold, resulting in lower and unknown false-positive probabilities. Our approach allows tests to be performed from time $t=1$ and configures adaptive thresholds to keep the false positive rate at the chosen level. }
    \label{fig:fig_1}
    \vspace{-6mm}
\end{figure}

Deploying a machine learning model to make decisions of real-world consequence carries risk. By evaluating the performance of the model on held out training instances one can obtain an unbiased estimate of how the model will perform on a stream of future deployment data. However, unbiasedness requires that the distribution underlying the deployment data remains identical to that which underlay the training data. In practice, not only can seemingly benign changes in the underlying process cause catastrophic deterioration in model performance, when feedback is delayed this can occur silently and damage can accumulate over time. Attempts to make machine learning models robust to such changes have so far had limited success \citep{taori2020measuring, ovadia2019can}. There is therefore demand for algorithms capable of detecting such changes directly and raising alerts such that adaptation or retraining processes can be triggered.

During deployment a model is passed features $x \in \mathcal{X}$ and tasked with predicting an associated unobservable label $y \in \mathcal{Y}$. The unobservability of labels makes it necessary to look for change in the feature space $\mathcal{X}$, which may have complex and high dimensional structure, such as for text or images. The resulting problem is often referred to as unsupervised drift detection and motivates the development of change detectors that are flexible enough to operate on various data domains.

Change detection algorithms contain threshold parameters that influence the frequency of false detections in the absence of change. The convention in the statistical literature is to set thresholds such that, in the absence of change, the false positive rate (FPR) at each time step is bounded from above. This is commonly referred to as \textbf{``controlling''} the false positive rate. However, the tightness of such bounds is usually completely unknown. Resulting detectors therefore operate at unknown false positive rates that can be -- and often are\footnote{This can often be verified directly from type-I error analyses in other works. It is sometimes even considered desirable.} -- orders of magnitude below the specified bound. 

We instead aim to set thresholds such that detectors operate at an actual FPR that is, with high approximation accuracy, equal to the desired FPR. We refer to this as instead \textbf{``targeting''} a desired false positive rate. Not only does this ensure that the significance of detections is known when they occur so that an appropriate response can be made, but it also ensures that statistical detection power is not hampered due to operating at some unknown false positive rate orders of magnitude below that which the practitioner is happy with. There exists few works that aim to target desired FPRs in this sense, particularly for the machine learning setting on which we focus where:
\vspace{-1mm}
\begin{itemize}
    \item random variables take values in a multivariate and potentially non-Euclidean domain,\vspace{-0.5mm}
    \item knowledge of the pre-change and post-change distributions is completely absent,\vspace{-0.5mm}
    \item there exists a large set of reference data from the pre-change distribution.\vspace{-0.5mm}
\end{itemize}
To the best of our knowledge the only existing work that tackles the same problem setting is that of \cite{li2019scan}, whose calibration process is motivated by asymptotic arguments which we later show can lead to significant miscalibration in practice.  Our contributions are therefore to: \vspace{-1mm}
\begin{enumerate}
    \item Present a novel, simulation-based and metric-agnostic approach to threshold setting that results in calibrated detectors where the false detection probabilities are known and kept constant across time steps.\vspace{-1mm}
    \item Show how an estimator of maximum mean discrepancy (MMD) can be used to define change detectors that can operate on streams of data residing in any domain on which a kernel can be defined.\vspace{-1mm}
    \item Show how to structure computations such that the cost of leveraging the minimum variance unbiased estimator of MMD is reduced from $O(N^2B)$ to $O(N^2+NB)$ during configuration and from $O(N^2)$ to $O(N)$ per timestep during operation, where $N$ and $B$ are the number of reference samples and bootstrap samples respectively.\vspace{-1mm}
    \item Make implementations available to use as part of the popular open-source Python library \texttt{alibi-detect} \citep{alibidetect}.
\end{enumerate}

\section{BACKGROUND}

\subsection{Problem Statement and Notation}\label{sec:prob}

Let $\bfZ=\{Z_t\}_{t \geq 1}$,\; $Z_t \in \mcz$,\; denote a stream of independent random variables generated by the process
\begin{equation}\label{eqn:model}
    Z_t \sim
        \begin{cases}
            p & \text{for $t < \tau$} \\
            q & \text{for $t \geq \tau$},
        \end{cases}
\end{equation}
where $p$ and $q$ respectively denote pre-change and post-change distributions over $\mathcal{Z}$ and $\tau \in \mathbb{N}^+ \cup \{\infty\}$ is an unknown change point.
We focus on the setting where both $p$ and $q$ are completely unknown, $\mathcal{Z}$ may be multivariate and non-Euclidean, and there exists a large set \mbox{$\tilde{\mathbf{z}}=\{\tilde{z}_i\}_{i=1}^N$} of i.i.d. reference instances from $p$.

To refer to windows of the data stream we use interval notation such that, for example, $\bfZ_{(t-W,t]}$ denotes the window of size $W$ starting at time $t-W+1$ and ending at time $t$. We use $\bfZ^{(J)}$ to denote a collection of $J$ random variables distributed i.i.d. according to $p$ and $\tbfZ^{(J)}$ to denote a random subsample of size $J$ from $\tilde{\mathbf{z}}$ where we will specify whether the sampling takes place with or without replacement.
We use dashes to distinguish random variables when necessary such that, for example, $\bfZ^{(J)}$ and $\bfZ'^{(J)}$ denote independent and identically distributed random variables. We use $\mathbb{E}_{\tau}$ and $P_{\tau}$ to denote expectations and probabilities corresponding to the stream with change point $\tau$.

Now consider the sequential change detection problem where at each time $t$ we test whether significant evidence exists to suggest that change has already occurred (i.e. whether $\tau \leq t$).  Letting $T$ denote the time at which a detection is made, we consider the problem of designing detection algorithms that minimise \cite{pollak1985optimal}'s formulation of the worst case expected detection delay
\begin{equation}
\text{EDD} = \sup_{q \neq p, \tau \geq 1} \mathbb{E}_{\tau}[T-\tau|T \geq \tau],    
\end{equation} 
subject to operating at a known expected runtime
\begin{equation}
    \text{ERT}=\mathbb{E}_{\infty}[T]
\end{equation} 
in the absence of change. Here expectations are over  $\mathbf{Z}$ and any stochasticity present in the detection algorithm\footnote{Note that this formulation of EDD differs from that of \cite{lorden1971procedures}, which would additionally take the supremum, rather than expectation, over $\{Z_t\}_{t=1}^{\tau-1}$.}, but not over $\tilde{\mathbf{z}}$ which is observed prior to configuring the algorithm. 

\subsection{Not All False Positive Rates Are Equal}
Note how the ERT corresponds to a rate of false positives. In the absence of change we expect a false positive every ERT time steps on average. However they do not necessarily occur at a constant rate. Assuming that a detector calibrated to $\text{ERT}=\mu$ does not make a false detection at times $1,...,t-1$, the probability of a false detection at time $t$ is not necessarily $1/\mu$. 

Detectors based on overlapping windows of size $W$ are usually incapable of making detections for $1 \leq t < W$, have disproportionately large probabilities of making false detections at times $W \leq t < 2W$ and then a disproportionately low probability thereafter. Detectors that instead let evidence accumulate indefinitely are more likely to make false detections later in the stream. These irregularities complicate the interpretation of the significance of detections when they occur. 

We therefore consider it desirable to not only target a specifiable $\text{ERT}=\mu$, but to do so in a manner such that false detections occur at a constant rate in the absence of change. In other words, conditional on $\tau=\infty$, the detection time $T$ should be distributed according to the geometric distribution $\text{Geom}(\alpha)$ with $\alpha=1/\mu$. Detectors then satisfy the \textit{memoryless} property \begin{equation}
    P_{\infty}(T-s > t|T>s) = P_{\infty}(T>t)\;\;\; \forall s,t \in \mathbb{Z}.
\end{equation} 

\subsection{Related Work}

Traditional approaches to sequential change detection either assume some degree of knowledge regarding the pre-change and post-change distributions \citep[e.g.][]{page1954continuous, lorden1971procedures} or are limited to the univariate case \citep[e.g.][]{kifer2004detecting, ross2012two, bifet2007learning}. Designing detectors that are flexible enough to detect any change in the distribution governing a multivariate data stream has been an area of recent focus \citep{bu2017incremental,li2019scan, liu2018accumulating,mozaffari2019online,hinder2020towards,chen2019sequential,kurt2020real}.

Change detectors typically work by (sometimes implicitly) testing for changes between a window of reference data and a window of `test' data. Whilst the reference window can be dynamic \citep[e.g.][]{chen2019sequential}, we focus on the setting where the reference window is fixed, as is most useful for detecting change from a distribution that generated a model's training set. One way to define test windows is in an adaptive manner where the window is allowed to grow whilst its contents is indicative of drift and reset to zero when no such indication is present. This strategy works well alongside complete \citep{page1954continuous} or partial \citep{lorden1971procedures, pollak1978optimality,chen2020high} knowledge of the pre-change and post-change distributions but otherwise either has an operation-time cost per time step that can grow unboundedly \citep{yu2020note} or considers incoming points only in isolation \citep{kurt2020real, mozaffari2019online, flynn2019change}. We therefore consider test windows of fixed size $W$ which sequentially receive the newest observation and release the $W$th oldest. This keeps the operation-time cost fixed and makes targeting a desired ERT tractable.

The inability of most change detectors to operate at (or even close to) a known ERT makes performance evaluations and comparisons difficult. Most commonly TP/TN/FP/FN rates are computed under various threshold values (corresponding to different, unknown, ERTs) and ROC/AUC-like metrics are compared. The degree to which detectors can be configured to operate with desired expected behaviour in the absence of change is rarely considered, despite its importance.

\subsection{B-statistic and LSDD-Inc}\label{sec:others}

In this section we describe two nonparametric and fixed window-size methods that consider change detection in the same setting as us, as described in Section~\ref{sec:prob}. At each time $t$, a test window $\twZ$ with unknown underlying distribution $q$ is considered and a test statistic $S_t=\hat{D}(\tbfz, \twZ)$ is computed as a sample-based estimate of a notion of distance $D(p,q)$ between the unknown distributions $p$ and $q$.

As the notion of distance \cite{bu2017incremental} use the least squares density difference (LSDD), defined as
$$D(p,q) = \int (p(z)-q(z))^2 \,dz.$$ \cite{li2019scan} instead use the maximum mean discrepancy (MMD) \citep{gretton2012kernel}, which can be defined in multiple ways but perhaps most simply as $$D_k(p,q) =  \sqrt{\mathbb{E}[k(X,X') + k(Y,Y') - 2k(X,Y)]},$$ for some kernel $k: \mcz \times \mcz \rightarrow \mathbb{R}$ and $X,X'\sim p$, $Y,Y' \sim q$. Crucially, both of these notions of distance admit two-sample estimators that can be incrementally updated at low cost. To estimate LSDD \cite{bu2017incremental} adopt a regularised estimator which involves fitting a Gaussian kernel model to the difference function $d(x)=p(x)-q(x)$. It can be updated in $O(1)$ time with respect to the number of reference samples $N$. To update an estimate of MMD at low cost, and to make (asymptotic) analysis of runtimes tractable, \cite{li2019scan} use the linear time\footnote{With respect to the size of the reference set $N$. We do not consider scaling with respect to the size of the test window $W \ll N$, chosen independently of $N$, important.} 
B-statistic. We later show, however, that the more powerful quadratic time estimator can be updated in linear time such that resorting to the linear time estimator is unnecessary.

Having defined a test statistic $\hat{D}: \mcz^N \times \mcz^W \rightarrow \mathbb{R}$ that can be updated at low cost we can monitor the test statistic trajectory $\{S_t\}_{t \geq 1}$. The difficult part is then to determine when the trajectory has deviated significantly from its expectation under the no-change null. The strategy adopted by both \cite{bu2017incremental} and \cite{li2019scan} is to detect change whenever $S_t > \hat{h}$ for some time-invariant threshold $\hat{h}$. \cite{li2019scan} define $\hat{h}$  to be an estimate of $h$, the unknown threshold that achieves a desired ERT. However their estimate is only accurate asymptotically (holds as $h \rightarrow \infty$) and leads to significant miscalibration in practice as we later show. \cite{bu2017incremental} take a simulation based approach which is more similar to ours, but control the ERT rather than target a desired level. Their approach is outlined in Algorithm~\ref{alg:lsddinc}, where superscripted pairs $(J,K)$ are used to denote quantities corresponding to a reference window of size $J$ and test window of size $K$.

\setlength{\textfloatsep}{8pt}
\RestyleAlgo{boxruled}
\LinesNumbered
\setlength{\algomargin}{3pt}
\begin{algorithm}[t]
  \small
  \caption{\fontsize{9.7}{11.64}\selectfont LSDD-Inc threshold configuration \label{alg:lsddinc}}
    \textbf{Input} reference set $\tbfz \in \mathcal{Z}^N$, window size $W$ and number of bootstrap samples $B$. \\
    \For{ $b=1,...,B$:}{
        Sample, \textit{with replacement} from $\tbfz$, a reference window $\tbfZ^{(W)}_b$ and test window $\tbfZ'^{(W)}_b$.\\
        Compute the corresponding estimate $S^{(W,W)}_b=\hat{D}(\tbfZ^{(W)}_b,\tbfZ'^{(W)}_b)$ of $\text{LSDD}(p,p)$.\\}
    Let $\hat{h}^{(W,W)}$ be the empirical $1-\alpha$ quantile of $\{S^{(W,W)}_b\}_{b=1}^B$.\\
    \textbf{Output} threshold $\hat{h} = \hat{h}^{(N,W)} := \mathbb{E}[\hat{D}(\bfZ^{(N)}, \bfZ^{(W)})] + (\hat{h}^{(W,W)}-\mathbb{E}[\hat{D}(\bfZ^{(W)}, \bfZ'^{(W)})])$, where the expectations have known values.
\end{algorithm}

Up until step 5 Algorithm~\ref{alg:lsddinc} is standard and estimates $h^{(W,W)}$, the $(1-\alpha)$-quantile of $\hat{D}(\bfZ^{(W)}, \bfZ'^{(W)})$, with an estimate $\hat{h}^{(W,W)}$ bootstrapped using with-replacement sampling from $\tbfz$. However step 6 is less standard and the justification is that the distribution of $\hat{D}(\bfZ^{(N)}, \bfZ^{(W)})$ will be more tightly centered on its expected value than that of $\hat{D}(\bfZ^{(W)}, \bfZ'^{(W)})$ and therefore $\hat{h}^{(N,W)}$ will correspond to a quantile of $\hat{D}(\bfZ^{(N)}, \bfZ^{(W)})$ that is greater than $1-\alpha$. This procedure therefore bounds the expected runtime from below rather than target it. The bound can be very loose if $N \gg W$ and the distribution of $\hat{D}(\bfZ^{(W)}, \bfZ'^{(W)})$ is much more diffuse than that of $\hat{D}(\bfZ^{(N)}, \bfZ^{(W)})$.  Additionally, and even more significantly, no correction is made to account for the bias resulting from the fact that consecutive test statistics are highly correlated.
In combination these biases result in an actual ERT that is often orders of magnitude greater than desired and therefore an EDD much larger than would be achieved with a detector targeting the desired ERT.

\section{CALM}\label{sec:calm}

Note that we are not in fact interested in the distribution of $\hat{D}(\bfZ^{(N)}, \bfZ^{(W)})$. Given that $\tbfz$ is known prior to both configuration and operation, we wish to configure the detector to operate at a desired ERT conditional on the particular realisation of the reference data that shall be used to compute test statistics -- not conditional on the reference data being drawn from the same underlying distribution. In other words, we are instead interested in the distribution of $\hat{D}(\tbfz, \bfZ^{(W)})$.

Ideally we would estimate the distribution of $\hat{D}(\tbfz, \bfZ^{(W)})$  by sampling $\bfZ^{(W)}_1,...,\bfZ^{(W)}_B$ from $p$. Without the ability to sample $p$, a naive bootstrap approach would be to instead use with-replacement samples $\tbfZ^{(W)}_1,...,\tbfZ^{(W)}_B$ from $\tbfz$. However this results in biasing quantile estimates of $\hat{D}(\tbfz, \bfZ^{(W)})$ downward due to the fact that the samples in the test window are also in the reference window; we refer to this as the \textit{window-sharing bias}.

This explains why \cite{li2019scan} instead use $\hat{D}(\bfZ^{(W)}, \bfZ'^{(W)})$ as an intermediary and target it using a standard bootstrap approach; assuming $W \ll N$ then the probability of shared samples between the reference and test windows, and therefore the resulting window-sharing bias, is very small. However, as mentioned earlier, corresponding thresholds can then only be used to lower-bound the ERT rather than target it. 

We instead propose using a reference window $\tbfZ^{(N-W)}$ sampled without replacement from $\tbfz$ during configuration and kept fixed throughout operation. Test statistics then take the form $S_t = \hat{D}(\tbfZ^{(N-W)},\twZ)$ and the quantity of interest is $h^{(N-W,W)}$, the $(1-\alpha)$-quantile of $\hat{D}(\tbfZ^{(N-W)},\bfZ^{(W)})$, which encompasses the randomness resulting from sampling the reference window. This quantity can then be estimated using the procedure described in Algorithm~\ref{alg:thresh_time_inv}. Within this procedure the process for sampling reference windows corresponds exactly to the process used during operation and does not introduce any bias. The process for sampling test windows corresponds to estimating the distribution of $\bfZ^{(W)}$ using that of $\tbfZ^{(W)}$ sampled without replacement from $\tbfz$. The fact that we are inferring a property of a function of $W$ random variables using a pool of size of $N$, where $N \gg W$, means that the bias introduced here is actually much less than that typically introduced by bootstrap approximations of functions of $N$ random variables\footnote{This is evidenced by the existence of the subsampling bootstrap and m-out-of-n bootstrap that approaches the $N=W$ case by first performing inference for a function of fewer variables and then applying corrections.}. For $W<N$, sampling without replacement (subsampling) is considered preferable and, crucially, allows the window-sharing bias to be completely eliminated by keeping the reference and test windows disjoint. Experiments evidencing this reduction of bias can be found in the supplementary material.

\setlength{\textfloatsep}{8pt}
\RestyleAlgo{boxruled}
\LinesNumbered
\setlength{\algomargin}{3pt}
\begin{algorithm}[t]
  \small
\caption{\fontsize{9.7}{11.64}\selectfont\mbox{Time-invariant threshold configuration}\label{alg:thresh_time_inv}}
\textbf{Input} reference set $\tbfz \in \mathcal{Z}^N$, window size $W$ and number of bootstrap samples $B$. \\
    \For{$b=1,...,B$:}{
        Sample, \textit{without replacement} from $\tbfz$, a reference window $\tbfZ^{(N-W)}_b$ of size $N-W$ and let the $W$ unsampled points $\bfU_b:= \tbfz \setminus \tbfZ^{(N-W)}_b$ act as a corresponding test window.\\
        Compute the corresponding estimate $S^{(N-W,W)}_b=\hat{D}(\tbfZ^{(N-W)}_b,\bfU_b)$ of $D(p,p)$.}
    Let $\hat{h}^{(N-W,W)}$ be the empirical $(1-\alpha)$-quantile of $\{S^{(N-W,W)}_b\}_{b=1}^B$.\\
    \textbf{Output} threshold $\hat{h}=\hat{h}^{(N-W,W)}$
\end{algorithm}

\subsection{Adjusting for Correlated Test Outcomes}\label{sec:adjusting}

The above process aims to configure $\hat{h}$ such that the first\footnote{Note how prior to time $t=W$ the test window is not full and a test statistic can therefore not be computed.}  test statistic $S_W$ satisfies $P(S_W>\hat{h})=\alpha$. However, the second test statistic $S_{W+1}$ is computed using almost the same data ($\bfZ_{(1,W+1]}$ rather than $\bfZ_{(0,W]}$) and is therefore highly correlated with $S_W$. Consequently, in the absence of change the probability $P(S_{W+1}>\hat{h} | S_W < \hat{h})$ of a false detection at time $t=W+1$ conditional on no false detection at time $t=W$ is some unknown value below $\alpha$. If we let $\alpha_t$ denote the false detection probability at time $t$ conditional on no prior false detections, even if $\hat{h}=h$ is perfectly estimated we have $\alpha_t<\alpha$ for all $t > W$ and therefore an expected runtime that takes some unknown value above that which was desired. We would instead like to estimate time-varying thresholds $h_t$ that result in $\alpha_t=\alpha$ for all $t$.

In estimating these time-varying thresholds we now consider sampling and fixing a reference window $\tbfZ^{(N-2W+1)}$ of slightly reduced size $N-2W+1$. At the first test time $t=W$ we wish to compare the test statistic $S_W = \hat{D}(\tbfZ^{(N-2W+1)},\bfZ_{(0,W]})$ against the threshold $h_W$ that satisfies $P(S_W > h_W) = \alpha$, which we can estimate using the bootstrap approach already described above. Then for $t>W$ we wish to compare $S_t$ against the threshold $h_t$ that satisfies
$$P\big(S_t > h_t |S_{t-1} \leq \hat{h}_{t-1} ,..., S_{W} \leq \hat{h}_W  \big) = \alpha.$$
This problem has been considered for adaptive window methods for which \cite{verdier2008adaptive} propose a solution for contexts where sampling from $p$ is possible. We now show how the same idea can be applied to the fixed window size context where we have only a finite sample $\tbfz$ from $p$. What makes this possible is the fact that when the window size is fixed we have
\begin{equation}
    h_{W} \neq h_{W+1} \neq ... \neq h_{2W-1} = h_{2W} = ...
\end{equation}
That is, the thresholds differ for the first $W$ tests but then remain constant. To see this note that for the first test at time $t=W$ the points in the test window are all unseen and have collectively contributed to 0 passed tests. For the second test we have $W-1$ points that have contributed to a passed test each and a new unseen point -- so the test is performed with a test window that has collectively contributed $W-1$ times to passed tests. For the third test we have $(W-1)+(W-2)$ passed test contributions as some have contributed to two passed tests. Up until the $W$-th test we receive more and more information that makes a passed test more likely. But then the information available for the $(W+1)$-th test (at time $2W$) is the same as that which was available for the $W$-th test -- every single test point has maximally contributed to passed tests (i.e. $(W-1)+(W-2)+...+1$ contributions). In other words, observing a passed test at time $t=2W-1$ doesn't introduce any information that causes $h_{2W}$ to differ from $h_{2W-1}$. 
The same principle applies thereafter, meaning that the problem is reduced to estimating the $W$ thresholds $h_{W},...,h_{2W-1}$. 

Like \cite{verdier2008adaptive} we simulate $B$ trajectories under the no-change null and sequentially estimate quantiles such that the trajectories used in the estimation of $h_t$ are only those which have remained below all previous thresholds (in our case $\hat{h}_{W},...,\hat{h}_{t-1}$), thereby enacting the conditioning. More concretely, defining $M=N-2W+1$ for compactness, our approach is described in Algorithm~\ref{alg:calm}. Note how the number of bootstrap samples remaining with which to compute quantile estimates gradually decreases such that approximately $B(1-\alpha)^W$ remain for the final estimate $\hat{h}_{2W-1}$ of $h_{2W-1}$. However, assuming an ERT of $\mu>W$ and recalling $\alpha=1/\mu$ we can note $(1-\alpha)^W > (1-\frac1W)^W \approx e^{-1}$ for all reasonable values of $W$ and therefore the number of bootstrap samples remains on the same order of magnitude throughout.

\setlength{\textfloatsep}{8pt}
\RestyleAlgo{boxruled}
\LinesNumbered
\setlength{\algomargin}{3pt}
\begin{algorithm}[t]
  \small
\caption{\fontsize{9.7}{11.64}\selectfont CALM threshold configuration}
\label{alg:calm}
\textbf{Input} reference set $\tbfz \in \mathcal{Z}^N$, window size $W$ and number of bootstrap samples $B$. \\
    \For{$b = 1,...,B$:}{
        Sample, \textit{without replacement} from $\tbfz$, a reference window $\tbfZ^{(M)}_b$ of size $M=N-2W+1$. \\
        Let $\bfU_b:=\tbfz \setminus \tbfZ^{(M)}_b$ act as a corresponding mini data stream of length $2W-1$.}
    Set $\mathcal{B}_W = \{1,2,...,B\}$. \\
    \For{$t=W,...,2W-1$:}{
        Compute  $\hat{h}_{t}^{(M,W)}$, the empirical $(1-\alpha)$-quantile  of $\{S^{(M,W)}_{t,b}\}_{b \in \mathcal{B}_t}$, where $S^{(M,W)}_{t,b}=\hat{D}(\tbfZ^{(M)}_b,(\bfU_b)_{(t-W,t]})$. \\
        Set $\hat{h}_t = \hat{h}_{t}^{(M,W)}$.\\
        Set $\mathcal{B}_{t+1} = \{b \in \mathcal{B}_t : S^{(M,W)}_{t,b} \leq \hat{h}_t\}$.}
    \textbf{Output} thresholds $(\hat{h}_W,...,\hat{h}_{2W-1})$.
\end{algorithm}

During operation $S_t=\hat{D}(\tbfZ^{(M)},\bfZ_{(t-W,t]})$ should be compared to $\hat{h}_t$ for $W \leq t < 2W-1$ and $\hat{h}_{2W-1}$ thereafter. We later show that this leads to detectors which, in the absence of change, have post-$W$ runtimes that accurately target the desired geometric distribution, i.e. $T-W \sim \text{Geom}(1/\mu)$. However, if desired, an adjustment can be made such that instead the absolute runtime $T$ targets the desired geometric distribution. In other words, we can start performing tests immediately at time $t=1$ rather than $t=W$. 

This can be achieved by initialising the detector with a test window of $W$ reference instances that were not sampled in the reference window $\tbfZ^{(M)}$. We think of this as prepending $W$ such instances to the stream as observations $Z_{-W+1},Z_{-W+2},...,Z_0$. In order for the first test outcome to depend on the extremity of $Z_1$ rather than that of $Z_{-W+1},Z_{-W+2},...,Z_0$ we check $S_0=\hat{D}(\tbfZ^{(M)}, \bfZ_{(-W,0]}) < h_W$ and resample if not. This enacts the conditioning necessary to allow us to compare $S_1=\hat{D}(\tbfZ^{(M)}, \bfZ_{(1-W,1]})$ to the conditional threshold $\hat{h}_{W+1}$. More generally $S_t$ should be compared to $\hat{h}_{W+t}$ for $1 \leq t < W$ and $\hat{h}_{2W-1}$ thereafter. 

Figure ~\ref{fig:fig_1} illustrates how our approach, which starts testing at time $t=1$ with adaptive thresholds, differs from approaches such as that of \cite{bu2016pdf} which starts testing at time $t=W$ with a fixed threshold that does not account for correlated test outcomes. For simplicity we here take $W=8$, let $\tilde{\mathbf{z}}$ contain $N=1000$ samples from $p=N(0,1)$ and define $S_t=\hat{D}(\tilde{\mathbf{z}},\bfZ_{(t-W,t]})$ where $\hat{D}$ simply returns the difference in sample means. We see that the threshold corresponding to the $1-\alpha=0.9$ quantile of the first test statistic corresponds to much higher quantiles at subsequent time steps. False positive rates, equal to the areas in red, only remain known and constant if thresholds are suitably adapted for the first $W$ tests.

\section{CALM-MMD}

The approach to threshold setting described in the previous section applies to any test statistic that takes the form of a two-sample estimator $\hat{D}(\tbfZ^{(M)}, \bfZ_{(t-W,t]})$ of distance between underlying distributions. In this section, owing to its flexibility to be applied to any domain on which a kernel can be defined, we focus more specifically on the case where $\hat{D}$ is an estimator of the squared maximum mean discrepancy $\text{MMD}_k^2(p,q)$ for some kernel $k: \mcz \times \mcz \rightarrow \mathbb{R}$.

The minimum variance unbiased estimator of this distance is given by the quadratic time estimator, which for samples $\mathbf{X} \in \mathcal{Z}^M$ from $p$ and $\mathbf{Y} \in \mathcal{Z}^W$ from $q$ is
\begin{equation}\label{eqn:mmd_hat}
\begin{aligned}
    \hat{D}(\mathbf{X}, \mathbf{Y}) = &\frac{\sum_{i\neq j} k(X_i,X_j)}{M(M-1)} + \frac{\sum_{i\neq j} k(Y_i,Y_j)}{W(W-1)} \\ &- \frac{\sum_{i,j} k(X_i,Y_j)}{MW}.
\end{aligned}
\end{equation}
Computing this from scratch has a cost of $O(N^2)$ and therefore a naive implementation of Algorithm \ref{alg:calm} during configuration would cost $O(N^2B)$. Moreover the cost during operation would be $O(N^2)$ per time step. However, we will now show that that these costs can be reduced to $O(N^2+NB)$ and $O(N)$ respectively. To keep notation compact we allow the kernel to operate on collections of observations such that $K=k(\tbfz, \tbfz)\in\mathbb{R}^{N \times N}$, for example, denotes the matrix with $(i,j)$-th entry $k(\tilde{z}_i, \tilde{z}_j)$. For a matrix $A$ we use shorthand notations $\sum_{i,j} A$ and $\sum_{i \neq j} A$ to denote the sum of all entries and off-diagonal entries respectively.

\subsection{Configuration}

During configuration, for each bootstrap $b=1,...,B$ and time $t=W,...,2W-1$ we are required to compute $S_{t,b}=\hat{D}(\tbfZ^{(M)}_b,(\bfU_b)_{(t-W,t]})$ where $\bfU_b=\tbfz \setminus \tbfZ^{(M)}_b$. Let \[
    K_b=
    \left[
    \begin{array}{c c}
    A_b & B_b \\
    B_b^{\top} & C_b
    \end{array}
    \right]
    \]
denote the kernel matrix $K$ with its rows and columns permuted such that $A_b=k(\tbfZ^{(M)}_b,\tbfZ^{(M)}_b)$, $B_b=k(\tbfZ^{(M)}_b, \bfU_b)$ and $C_b=k(\bfU_b,\bfU_b)$. Further define the submatrix $B_{t,b}=k(\tbfZ^{(M)}_b, (\bfU_b)_{(t-W,t]})$ of $B_b$ and $C_{t,b}=k((\bfU_b)_{(t-W,t]},(\bfU_b)_{(t-W,t]})$ of $C_b$. We can then write
\begin{equation}
    S_{t,b} = \frac{\sum_{i \neq j}A_b}{M(M-1)}+\frac{\sum_{i\neq j}C_{t,b}}{W(W-1)}-2\frac{\sum_{i,j}B_{t,b}}{MW}.
\end{equation}
By noting the relation $\sum_{i \neq j}K = \sum_{i \neq j}K_b = \sum_{i \neq j}A_b + \sum_{i \neq j}C_b + 2\sum_{i, j}B_b$ and amortizing the cost of initially computing $\sum_{i \neq j}K$ across all bootstrap samples we can compute $S_{t,b}$ for all $t=W,...,2W-1$ using only the submatrices $B_b$ and $C_b$. This results in a cost per bootstrap sample of $O(N)$. In addition to the single $O(N^2)$ cost of computing  $\sum_{i \neq j}K$, this results in a total configuration cost of $O(N^2+NB)$.

\subsection{Operation}

During operation we have a fixed reference window $\tbfZ^{(M)}$ and at each time step need to compute $S_t=\hat{D}(\tbfZ^{(M)}, \bfZ_{(t-W,t]})$. Let \[
    K_t=
    \left[
    \begin{array}{c c}
    K' & B_t \\
    B_t^{\top} & C_t
    \end{array}
    \right]
    \]
where $K'=k(\tbfZ^{(M)},\tbfZ^{(M)})$, $B_t=k(\tbfZ^{(M)},\bfZ_{(t-W,t]})$ and $C_t=k(\bfZ_{(t-W,t]},\bfZ_{(t-W,t]})$. We can then write
\begin{equation}
    S_{t} = \frac{\sum_{i \neq j}K'}{M(M-1)}+\frac{\sum_{i\neq j}C_{t}}{W(W-1)}-2\frac{\sum_{i,j}B_{t}}{MW}.
\end{equation}
We can compute $\sum_{i \neq j}K'$ just once during the configuration phase and therefore update the test statistic at cost $O(N)$ during operation. Although we have focused on scaling with respect to $N$ rather than $W \ll N$ we additionally note that caching certain computations (similarly to \cite{li2019scan}) allows $\sum_{i,j}B_{t+1}$ to be computed from $\sum_{i,j}B_{t}$ in $O(N)$ time (rather than $O(NW)$) and $\sum_{i\neq j}C_{t+1}$ to be computed from $\sum_{i\neq j}C_{t}$ in $O(W)$ time (rather than $O(W^2)$).

\section{EXPERIMENTS}

\subsection{Calibration and Power Comparison}

We compare, across a range of desired expected runtimes and pre-change and post-change distributions, the performance of the proposed change detectors with the MMD-based B-statistic detector of \cite{li2019scan} and LSDD-Inc detector of \cite{bu2017incremental}. Focus is on comparing the degree to which detectors can be calibrated to operate with known expected behaviour, or in other words, how closely average runtimes (ARTs) match desired expected runtimes (ERTs). However we additionally explore whether calibration comes at the expense of detection power, or in other words, the ability to respond to change with short delay.

\subsubsection{Experimental setup}

For these comparisons we generated data synthetically from known pre-change and post-change distributions. We chose the following 4 combinations, visualised in the supplementary material, with the first two taken directly from \cite{li2019scan} and the second two chosen to be similar in nature to D3-D6 in \cite{bu2017incremental} but adapted to be less Gaussian to provide diversity from the first two.
\vspace{-1mm}
\begin{description}
\vspace{-2mm}
  \item[D1] (Gaussian mean shift): Pre-change distribution of $\mathcal{N}(\textbf{0}, I_{20})$ and post-change distribution of $\mathcal{N}(0.3 \textbf{1}, I_{20})$, where $\textbf{1}$ is a vector of ones.
  \vspace{-1mm}
  \item[D2] (Gaussian covariance change): Pre-change distribution of $\mathcal{N}(\textbf{0}, I_{20})$ and post-change distribution of $\mathcal{N}(\textbf{0}, \Sigma)$ with diagonal covariance matrix $\Sigma$ satisfying $\Sigma_{ii}=1$ for $1 \leq i \leq 10$ and $\Sigma_{ii}=2$ for $11 \leq i \leq 20$.
  \vspace{-1mm}
  \item[D3] (Uniform square to uniform diamond): Pre-change distribution is the uniform distribution on the two dimensional square with coordinates $(\pm 1, \pm 1)$. Post-change distribution is the uniform distribution on the diamond with coordinates $(0,2)$, $(2,0)$, $(-2,0)$ and $(0,-2)$.
  \vspace{-1mm}
  \item[D4] (Hollowing of uniform square): Pre-change distribution is the uniform distribution on the two dimensional square with coordinates $(\pm 1, \pm 1)$. Post-change distribution is the uniform distribution on the same square but with the inner square with coordinates $(\pm 1/2, \pm 1/2)$ removed.
\end{description}
\vspace{-1mm}

To allow comparison to the B-statistic and LSDD-Inc detectors which can not detect until the test window is full, we did not use the modification described at the end of Section \ref{sec:adjusting} and focused on the distribution of the post-$W$ runtime. To obtain worst case average detection delays we used a change point of $\tau=W+1$ such that the detector is required to respond starting from a test window full of data from the pre-change distribution. To report average runtimes in the absence of change the post-change distribution was set to the pre-change distribution. The results are reported as averages over 100 configurations (each with different reference sets) and 500 runtimes per configuration, resulting in averages over 50000 runtimes.

For the B-statistic detector we configured thresholds as in the paper using the recommended skewness correction. For the LSDD-Inc detector, as discussed at the end of Section \ref{sec:others}, we found that not accounting for correlation between test outcomes and estimating thresholds corresponding to a reference set of size $N$ with bootstraps using reference sets of size $W$, resulted in average runtimes orders of magnitude greater than the specified lower bound. Therefore in order for experiments to complete and comparable results to be obtained we subsampled reference sets of size $W$ during the operational phase in order to correspond to the size of the reference sets used in the bootstrapping procedure. We compared these two detectors to CALM-MMD and CALM-LSDD using the same distance estimators but within the framework set out in Section \ref{sec:calm}. For the MMD-based detectors we use the Gaussian radial basis function (RBF) $k(z,z')=\exp(-||z-z'||_2^2/2\sigma^2)$ as the kernel using the popular median heuristic \citep{gretton2012kernel} where $\sigma$ is taken to be the median of the pairwise distances between reference samples.
\subsubsection{Results}
\setlength{\tabcolsep}{3pt}
\renewcommand{\arraystretch}{1.2}
\begin{table}
    \centering
    \caption{Comparison of miscalibrations and reductions on problems D1-D4, averaged over expected runtimes of 128, 256, 512 and 1024. Miscalibration is defined as the relative difference between average and expected runtimes in the absence of change. Reduction is defined as the relative reduction in detection time that results from a change occurring.
    \vspace{-0.1cm}
    }\label{tab:results}
    \scriptsize
        \begin{tabular}{| *{7}{c|} }
            \hline
            Method & \multicolumn{2}{c|}{Miscalibration} & \multicolumn{4}{c|}{Reduction} \\ 
         \hline
         & $\text{D1 \& D2}$ & $\text{D3 \& D4}$ & $\text{D1}$ & $\text{D2}$ & $\text{D3}$ & $\text{D4}$\\
         \hline
         $\text{B-statistic}$ & 0.253 & 0.201 & \textbf{0.951} & \textbf{0.910} & \textbf{0.903} & \textbf{0.562} \\
         $\text{CALM-MMD}$ & \textbf{0.010} & \textbf{0.010} & \textbf{0.951} & 0.909 & \textbf{0.903} & 0.560 \\
         \hdashline
         $\text{LSDD-Inc}$ & 0.202 & 0.187 & 0.936 & 0.866 & 0.888 & 0.522 \\
         $\text{CALM-LSDD}$ & \textbf{0.010} & \textbf{0.014} & \textbf{0.950} & \textbf{0.921} & \textbf{0.933} & \textbf{0.700} \\
        \hline
    \end{tabular}
    \end{table}
    
  \begin{figure}
    \begin{subfigure}{0.495\linewidth}
      \includegraphics[trim={8 0 45 40}, clip, width=\linewidth]{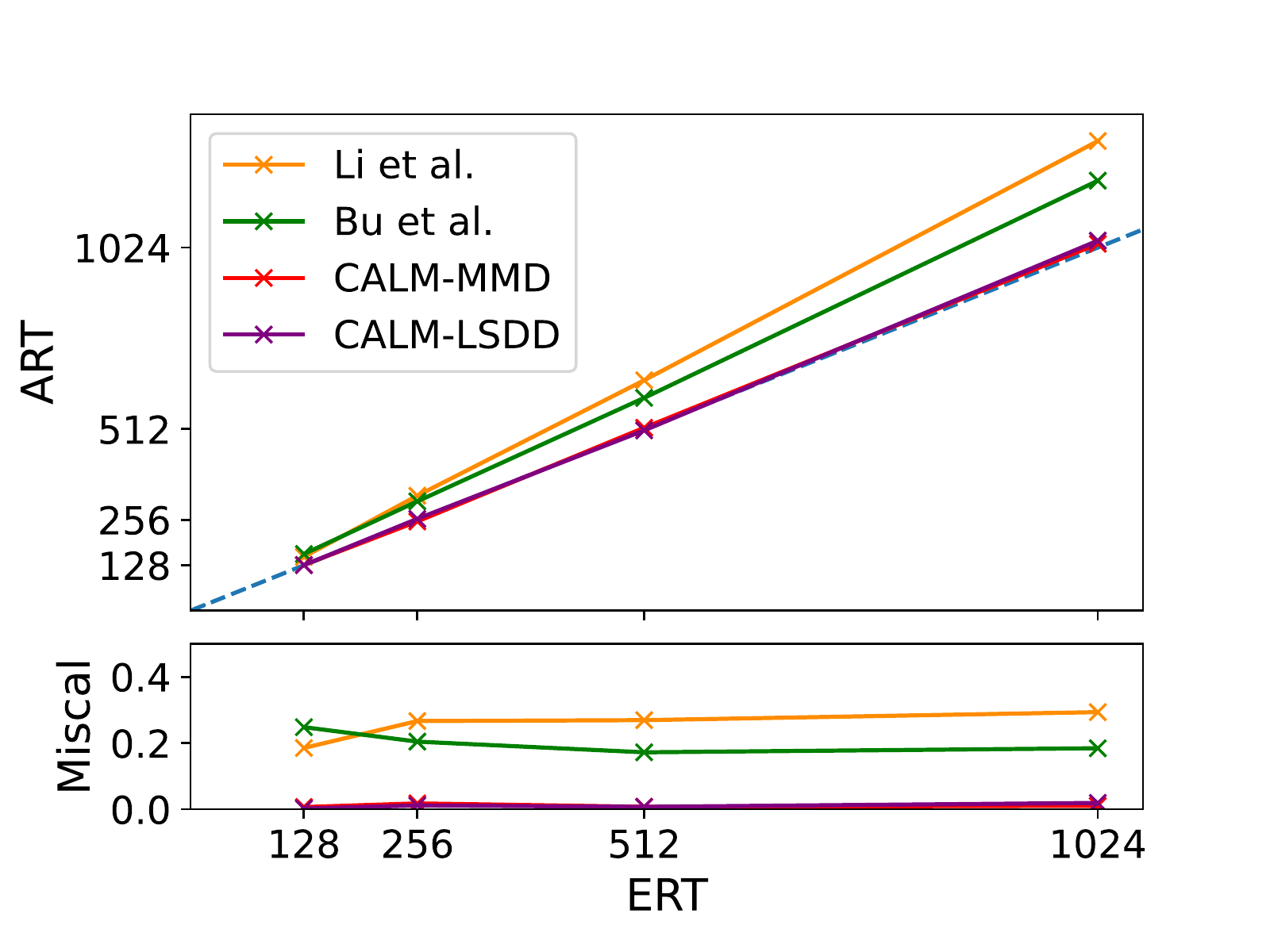}
      \caption{\small{Gaussian pre-change distribution (D1 \& D2)}}
      \label{fig:arts_a}
    \end{subfigure}
    \begin{subfigure}{0.495\linewidth}
      \centering
      \includegraphics[trim={8 0 45 40}, clip, width=\linewidth]{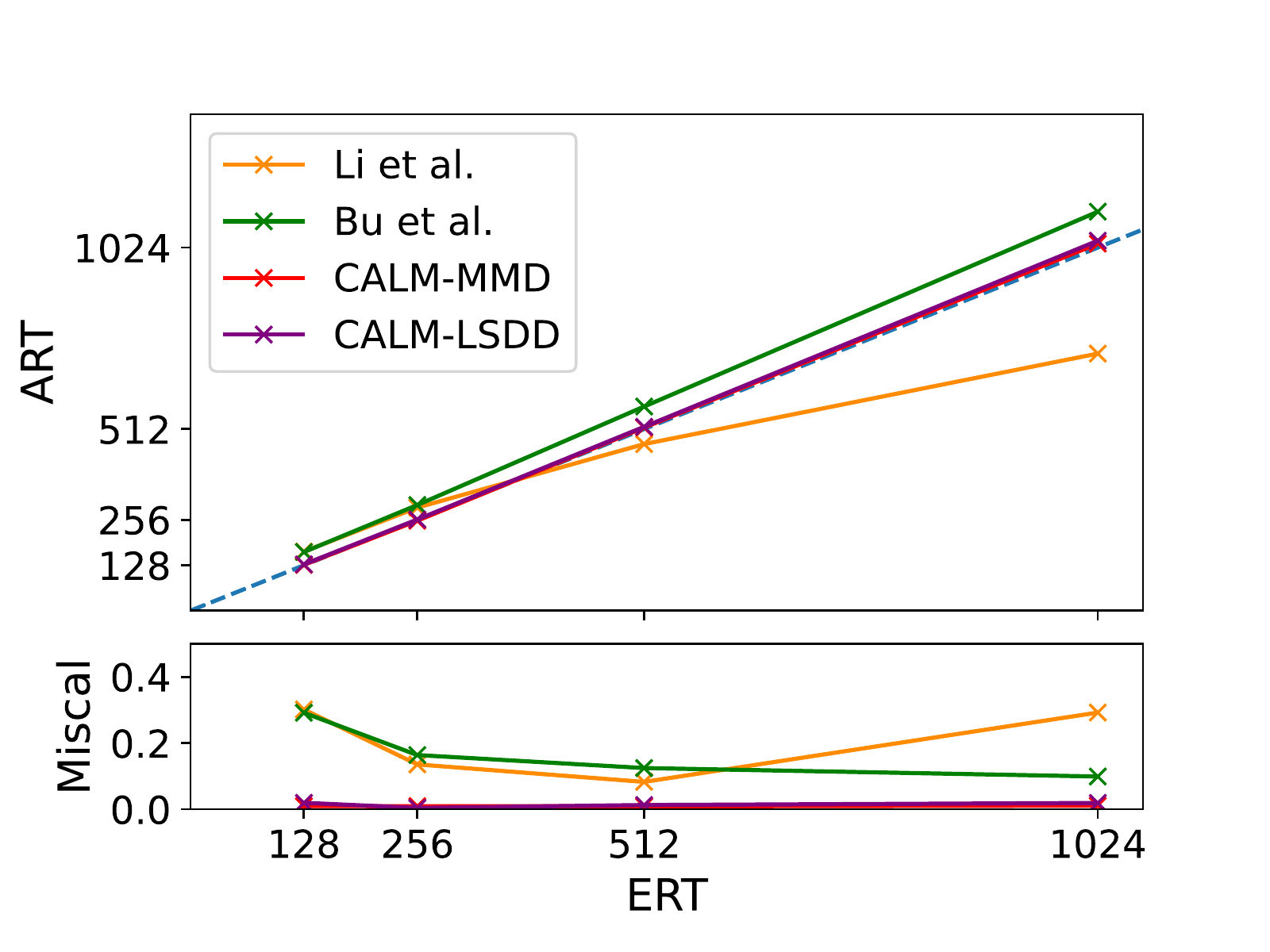}
      \caption{\small{Uniform pre-change distribution (D3 \& D4)}}
      \label{fig:arts_b}
    \end{subfigure}
    \caption{Desired expected runtimes (ERTs) plotted against actual average runtimes (ARTs) for each detector and pre-change distribution, with corresponding miscalibration metrics plotted beneath.}
    \label{fig:arts}
  \end{figure}
  For this comparison we fixed the size of the reference set to $N=1000$, the window size to $W=25$ and used $B=25000$ bootstrap samples to obtain threshold estimates. To compare calibration, for each detector we report the average runtime (ART) in the absence of change under configurations targeting ERTs of 128, 256, 512, and 1024. When plotted as a graph of ART vs ERT a well calibrated detector should lie close to the diagonal. We consider the relative error $|\text{ART}-\text{ERT}|/ \text{ERT}$ as a metric of miscalibration, where lower values represent better calibrated detectors. 
  
  Figure \ref{fig:arts} visualises the miscalibration of the four detectors for both pre-change distributions and across the four specified ERTs, the average across which is recorded in Table \ref{tab:results}. We see that the B-statistic and LSDD-Inc detectors typically operate at expected runtimes that deviate from those desired by between 15\% and 25\%. Moreover, the miscalibration varies depending on the problem and desired ERT. For the Gaussian pre-change distribution the miscalibration leads to average runtimes which are 20-25\% higher than desired across a range of ERTs. However on the uniform square the B-statistic detector operated at around 30\% above the desired level when an ERT of 128 is desired but around 30\% below the desired level when an ERT of 1024 is desired. This suggests that practitioners deploying such a detector should be aware that the actual false positive rate could be anywhere between 30\% below and 30\% above the level they desired.

\begin{figure}
  \includegraphics[trim={22 0 7 0}, clip, width=\linewidth]{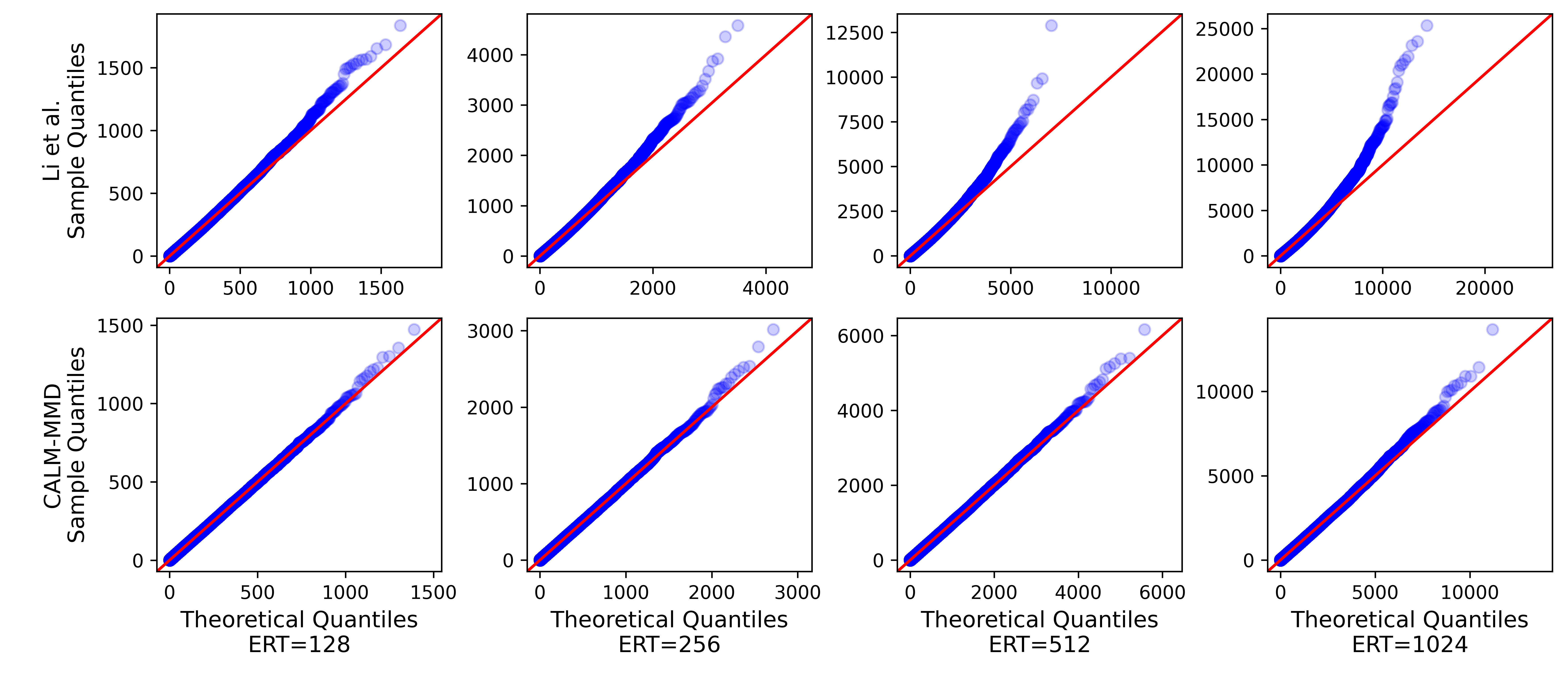}
    \caption{Q-Q plots of 50000 runtimes obtained in the absence of change under various expected runtimes, with theoretical quantiles corresponding to the geometric distribution with mean equal to the sample average.}
    \label{fig:qq}
\end{figure}

  By contrast, the detectors configured using the CALM methodology invariably operate at ERTs within approximately 1\% of the desired level, representing a 20$\times$ reduction in miscalibration. Moreover, the CALM methodology not only accurately targets the mean of the distribution of runtimes in the absence of change, but the entire distribution. The Q-Q plots in Figure~\ref{fig:qq} confirm that the distribution of runtimes is memoryless, with only a handful of the most extreme of the 50000 points lying away from the diagonal in each plot. By contrast using MMD within the framework of \cite{li2019scan} results in detectors with some unknown runtime distribution where short runtimes are underrepresented and large runtimes overrepresented relative to the corresponding memoryless distribution. The Q-Q plots are adjusted for the ARTs that are actually obtained such that deviation from the diagonal represents deviation from the best fitting geometric distribution, not that with mean equal to the ERT. We are not doubly penalising miscalibration of the mean.

\begin{figure}[t]
    \centering
        \includegraphics[trim={0 0 0 5}, clip, width=0.9\linewidth]{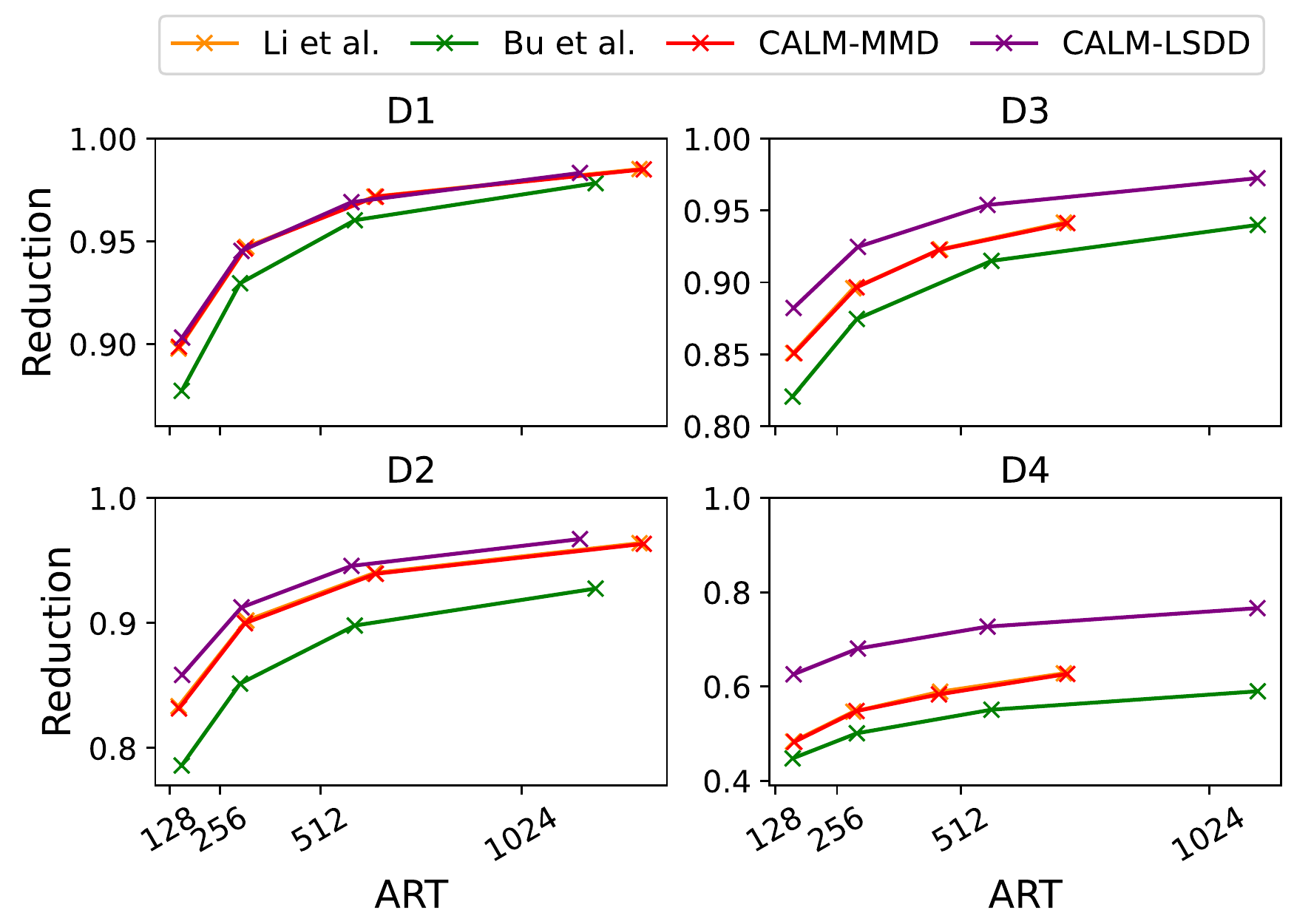}
        \caption{Plots of average runtimes in the absence of change against relative reductions resulting from change (ART-ADD/ART), where higher reductions represent more powerful detectors. Note that the red and orange curves align almost exactly.}
        \label{fig:adds}
  \end{figure}

We now consider whether this improved calibration comes at the cost of less powerful detectors. However, comparing the power of detectors operating at different ERTs is difficult. Merely comparing average detection delays (ADDs) favours detectors operating at lower ERTs and therefore miscalibrated detectors operating above the desired ERT get doubly penalised. To provide a fair comparison we performed additional experiments where we reconfigured the CALM-MMD and CALM-LSDD detectors to operate at the ERTs achieved by the corresponding detectors of \cite{li2019scan} and \cite{bu2017incremental}. We then recorded new ADDs to allow direct comparisons between detectors using the same notion of distance and operating at the same ERT. As a notion of power we report the relative reduction in runtime that results when a change occurs, i.e. $(\text{ART}-\text{ADD})/\text{ART}$, such that a perfect detector would score a value of 1 and a useless detector a value of 0. 
Figure \ref{fig:adds} allows, for each dataset D1-4, the power of the detectors to be compared. The average reduction across ERTs is recorded in Table \ref{tab:results}.

At least for the problems considered, there is no discernible difference in power between the B-statistic detector and CALM-MMD, with average reductions matching to two decimal places on all four problems, as shown by the overlapping curves in Figure~\ref{fig:adds}. This is despite the latter using (at linear time cost during operation) the quadratic time estimator of MMD. The difference between the LSDD-Inc detector and CALM-LSDD is much greater, which is unsurprising due to having to subsample the reference data as previously described. Although we should be careful when comparing detectors operating at different ERTs, comparison of the full curves in Figure~\ref{fig:adds} show that within the CALM framework the LSDD estimator resulted in more powerful detectors than the MMD estimator for these problems.
\subsection{Medical Imaging Example}
To demonstrate applicability to unsupervised drift detection problems of practical interest we now consider the Camelyon17-WILDS dataset \citep{koh2020wilds} containing tissue scans for classification as benign or cancerous. \cite{koh2020wilds} show that models trained to classify scans from a mixture of three hospitals suffer a performance drop when tasked with classifying scans from an unseen fourth hospital with a subtly different underlying distribution. It is therefore important to detect when such differences arise in order to prevent avoidable misdiagnoses. Labels (benign/cancerous) are assumed to be unavailable at detection time, making it necessary to detect change in the distribution of features (scans).

We took the distribution underlying scans from the three training hospitals to be our pre-change distribution $p$ and the distribution underlying scans from the unseen fourth hospital to be our post-change distribution $q$.
 We scaled up to $N=20000$, $\text{ERT}=5000$, $W=100$ and $B=100000$ and used the more expensive CALM-MMD detector. We trained an autoencoder to project the $96 \times 96 \times 3$ scans onto 32-dimensional vectors; see the supplementary material for details.  We configured the detector 150 times (each with newly trained autoencoders) and obtained 100 runtimes per configuration, resulting in a total of 15000 runtimes over which to average.

\begin{figure}
    \centering
        \includegraphics[trim={5 15 33 22}, clip, width=0.52\linewidth]{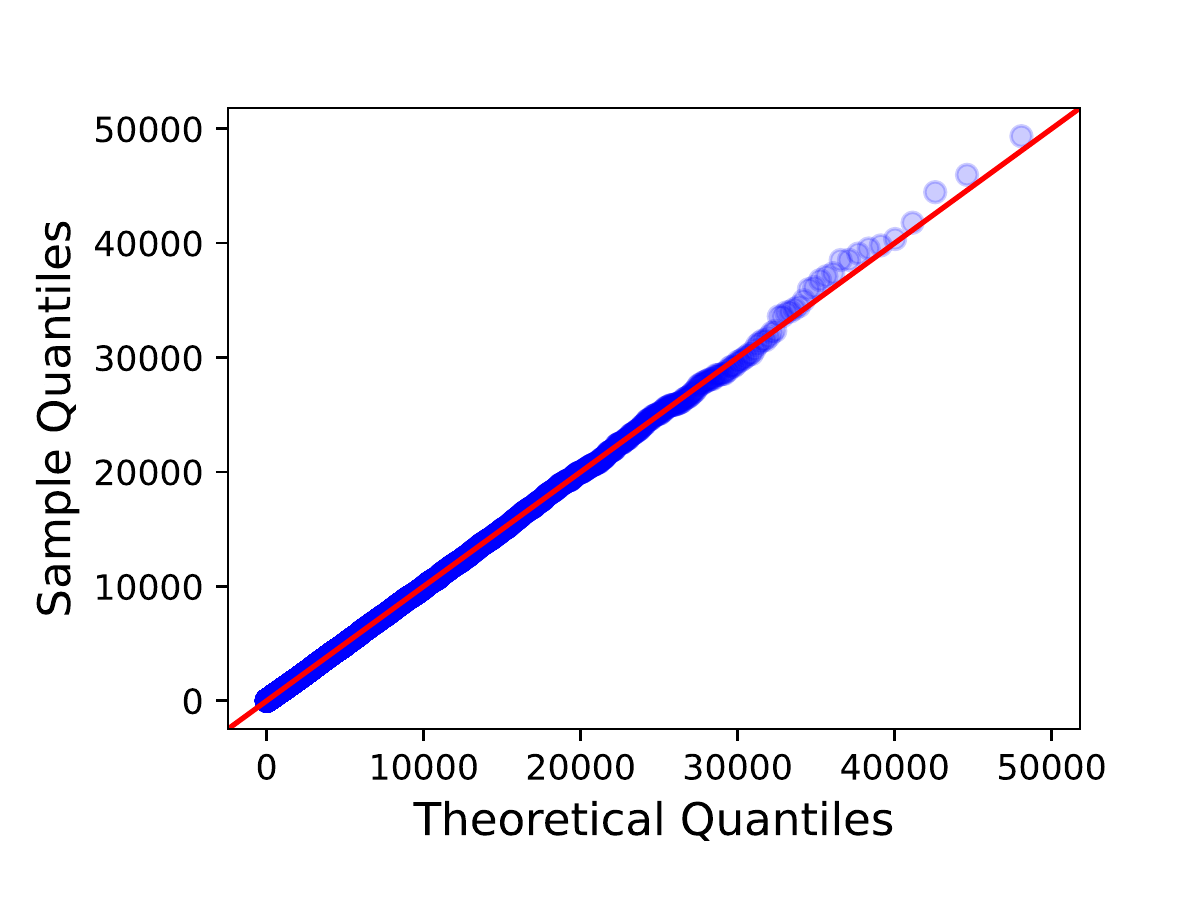}
        \caption{Q-Q plot of the runtimes obtained in the absence of change on Camelyon17 data. Theoretical quantiles correspond to the Geometric(5000) distribution.}
        \label{fig:camqq}
\end{figure}
 In the absence of change the average runtime was 4949, representing a miscalibration of 1.0\%. The Q-Q plot in Figure~\ref{fig:camqq} shows that this was achieved with runtimes closely following the desired memoryless distribution. By contrast the average detection delay following change was 52.8, representing a relative reduction of 98.9\% from the ART in the absence of change.

\section{CONCLUSION}
In this paper we focused on the problem of automatically configuring thresholds of change detection algorithms such that, in the absence of change, the distribution of runtimes follows the geometric distribution with the desired mean. We described a metric agnostic method for achieving this property and show that when used with the two-sample estimators of MMD and LSDD adopted by recent works the resulting detectors can then operate with known behaviour without compromising on responsiveness to change.

\subsubsection*{Acknowledgements}
We would like to thank Ashley Scillitoe for his help integrating our research into the open-source Python library \texttt{alibi-detect}. We are also grateful to the anonymous reviewers for their comments and suggestions.

\bibliography{main}

\clearpage
\appendix
\thispagestyle{empty}
\onecolumn \makesupplementtitle
\section{ADDITIONAL EXPERIMENTS: WINDOW-SHARING BIAS}

To demonstrate the existence and elimination of the window-sharing bias discussed in Section 3 we consider a simple example where the reference distribution $p$ is the $d$-dimensional isotropic Gaussian $N(\mathbf{0}, I_d)$, we have a reference set $\tbfz$ of size $N$ and we wish to estimate the distribution of $\hat{D}(\tbfZ^{(N-W)}_b,\bfZ^{(W)}_b)$ for some two-sample distance estimator $\hat{D}: \mcz^{N-W} \times \mcz^W \rightarrow \mathbb{R}$. We let $N=1000$, $W=25$ and consider the estimators of both MMD and LSDD. 

We are interested in comparing the accuracy of bootstrap estimates using with-replacement and without-replacement sampling. For the former the distribution of $\hat{D}(\tbfZ^{(N-W)},\bfZ^{(W)})$ , with $\tbfZ^{(N-W)}$ sampled with replacement from $\tbfz$, is estimated using bootstrap samples from $\tbfz$ where both the reference and test windows are sampled with-replacement. For the latter the distribution of $\hat{D}(\tbfZ^{(N-W)},\bfZ^{(W)})$, with $\tbfZ^{(N-W)}$ sampled without replacement from $\tbfz$, is estimated using bootstrap samples from $\tbfz$ where both the reference and test windows are sampled without replacement such that they are disjoint. For both cases we compare the empirical distribution obtained using 25000 bootstrap samples with an empirical distribution formed from 25000 samples from the true distribution being targeted, where the test window $\bfZ^{(W)}$ is sampled from $N(\mathbf{0}, I_d)$.

\begin{figure}[b!]
    \vspace{-5mm}
    \begin{subfigure}{0.35\textwidth}
      \includegraphics[trim={0 5mm 0 -19mm},clip, width=\textwidth]{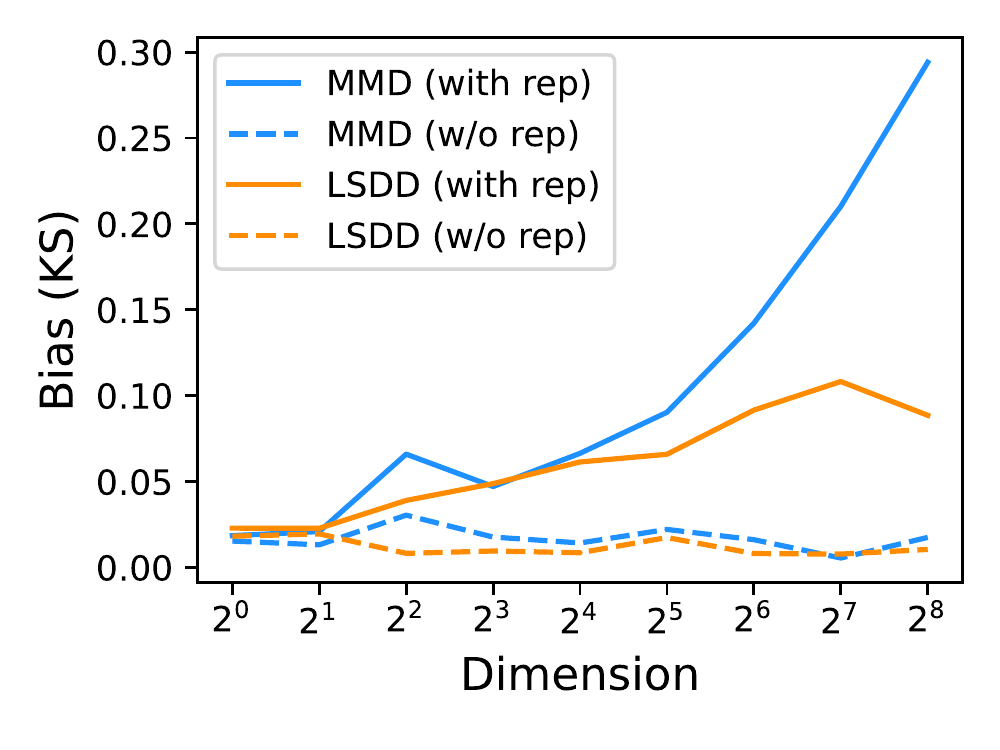}
      \caption{Kolmogorov-Smirnov distance between bootstrap samples and a sample from the distribution being targeted, plotted against the dimension of the Gaussian.}
      \label{fig:ws_bias_a}
    \end{subfigure}
    \hspace{0.5cm}
    \begin{subfigure}{0.6\textwidth}
    \vspace{3mm}
      \includegraphics[trim={13mm 0 15mm 0},clip, width=\textwidth]{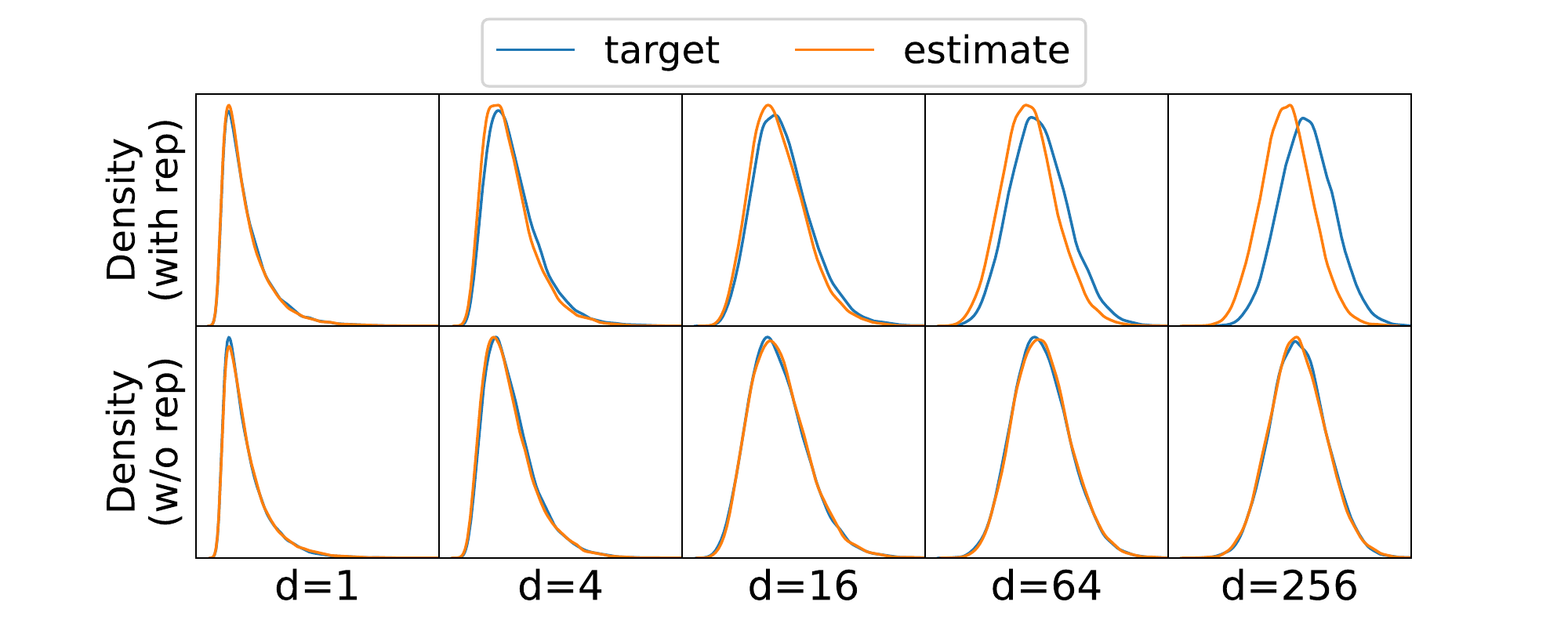}
      \caption{Probability density functions of the bootstrap samples and a sample from the distribution being targeted, corresponding to the MMD distance and a range of dimensions for the Gaussian.}
      \label{fig:ws_bias_b}
    \end{subfigure}

    \caption{A comparison of the accuracy of with-replacement and without-replacement bootstrap approximations of the distribution of two-sample distance estimators. Estimators of both MMD and LSDD are considered and both reference and test samples are drawn from a $d$-dimensional Gaussian.}

    \label{fig:ws_bias}
\end{figure}

Figure \ref{fig:ws_bias_b} shows the targeted and estimated distributions for a range of dimensions for the MMD case. We see that in the univariate setting both with-replacement and without-replacement sampling fares well and the distributions align. However as the dimension increases the alignment noticeably deteriorates when sampling with replacement as window sharing exerts a downwards bias on the estimates. This is unsurprising because MMD is based on pairwise similarities and in high dimensions fewer pairs of samples are similar to each other; therefore the bias introduced by the high similarity between multiple draws of the same instance is higher. Figure \ref{fig:ws_bias_a}, which plots the Kolmogorov-Smirnov two-sample distance between the targeted and estimated distributions against dimension, demonstrates this effect is true also for the LSDD distance (albeit to a lesser extent). The increased accuracy with which the distribution of without-replacement bootstrap samples approximates the distribution of interest means that quantile estimates are more accurate and therefore desired ERTs more accurately targeted.

\section{MAIN EXPERIMENTS: ADDITIONAL DETAILS}

\subsection{Calibration and Power Comparison}

\begin{figure}[t!]
  \centering
  \includegraphics[width=0.7\linewidth]{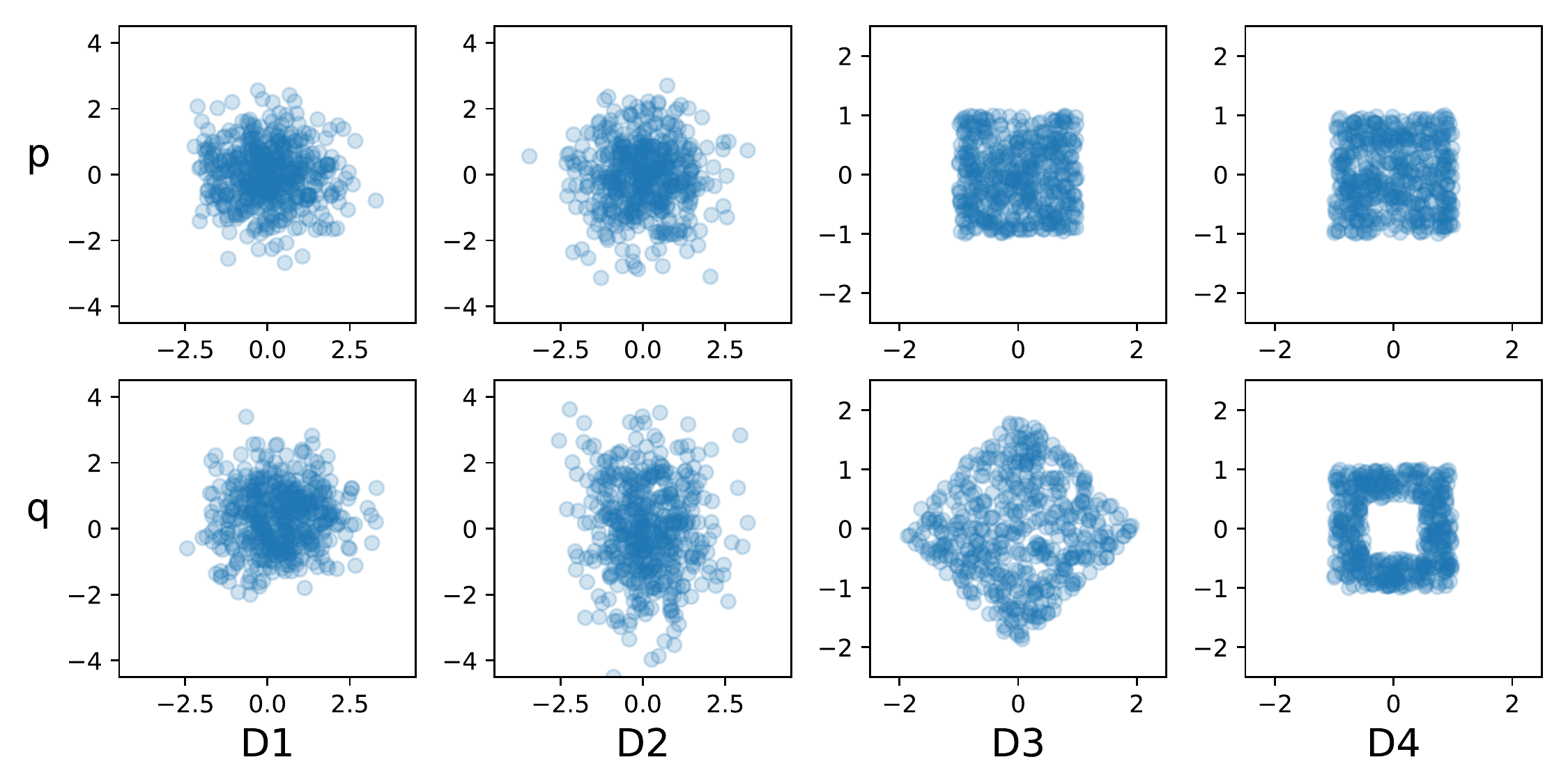}
  \caption{Plots of samples from the pre-change ($p$) and post-change ($q$) distributions for the four problems D1-D4. For D1 and D2 the plots corresponds to the 2-dimensional equivalent of the 20-dimensional problem under consideration.}
  \label{fig:toy_data}
\end{figure}

Plotted in Figure~\ref{fig:toy_data} are samples from the pre-change and post-change distributions corresponding to problems D1-D4 used for the calibration and power comparisons in Section 5.1.

\subsection{Medical Imaging Example}
\begin{figure}[b!]
    \centering
      \includegraphics[trim={40 45 40 82}, clip, width=0.62\linewidth]{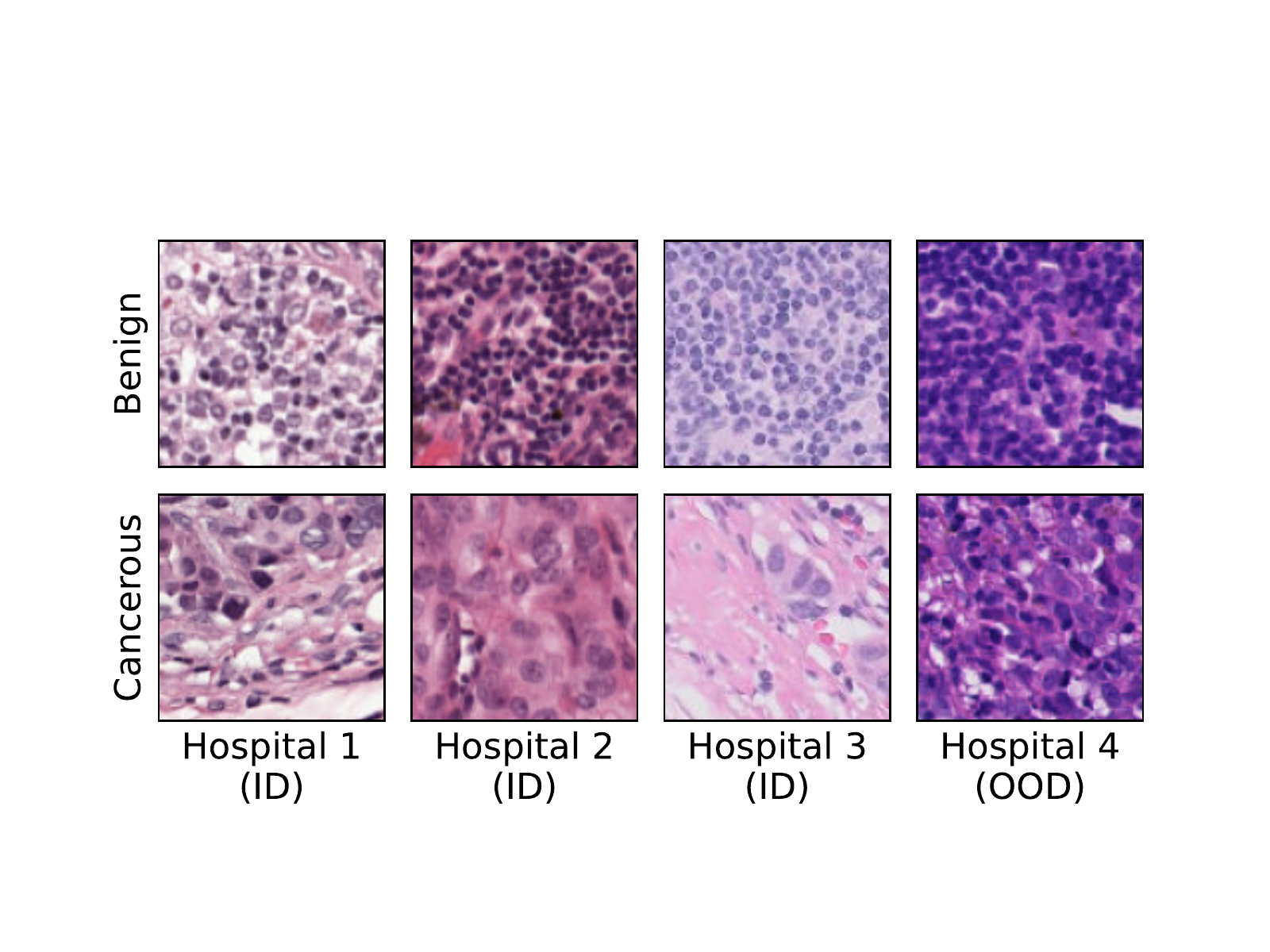}
      \caption{Scans from four hospitals contained in the Camelyon17-WILDS dataset. The first three are from the hospitals included in the training data whereas the fourth is from a hospital not represented in the training data.}
      \label{fig:camelyon}
\end{figure}

Pictured in Figure~\ref{fig:camelyon} are samples from four hospitals contained in the Camelyon17-WILDS dataset. The training data distribution, which we use as our pre-change distribution $p$, contains samples from the first three whereas the test data distribution, which we use as our post-change distribution $q$, contains samples from a fourth hospital not included in the training data. 

 Adhering to what we consider a recommended workflow in settings where the raw data is unstructured and of dimensionality likely far large than the underlying intrinsic dimensionality, we performed a preprocessing step to project the $96 \times 96 \times 3$ image patches onto a more structured lower dimensional representation.

To do this we trained an autoencoder to reconstruct the patches whilst passing them through a lower dimensional space of dimension $d=32$. For the encoder we used five convolutional layers, each separated by ReLU nonlinearities, which gradually reduce the spatial dimension from $96\times96$ to $1\times1$ and increase the number of channels from $3$ to $32$. The decoder is of symmetric form, mapping the 32 dimensional encoding vector back onto a $96 \times 96 \times 3$ image. Crucially, the autoencoder was trained using a split of the data which then no longer served as part of the reference set. We split the data such that half (10000) of the instances were used to learn the representation and the other half were used for testing. The autoencoder was trained using the Adam optimizer with a learning rate of 0.001 on batches of size $32$ for $25$ epochs. The change detectors were then applied to the 32-dimensional vectors that resulted from passing the image patches through the trained encoder.

\end{document}